\documentclass[aps,prd,preprint,groupedaddress]{revtex4-1}

\begin{document}

\title{ Decay constants in soft wall AdS/QCD revisited  }



\author{Nelson R. F. Braga}
\affiliation{Instituto de F\'{\i}sica,
Universidade Federal do Rio de Janeiro, Caixa Postal 68528, RJ
21941-972 -- Brazil}

\author{ M. A. Martin Contreras}
\affiliation{ High Energy Group, Department of Physics, Universidad de los Andes, Carrera 1, No 18A - 10, Bloque Ip, ZIP 111711, Bogot\'{a}, Colombia}

\author{Saulo Diles}
\affiliation{Instituto de F\'{\i}sica,
Universidade Federal do Rio de Janeiro, Caixa Postal 68528, RJ
21941-972 -- Brazil}


\begin{abstract}
Phenomenological AdS/QCD models, like hard wall and soft wall, 
provide  hadronic mass spectra in  reasonable consistency with experimental and (or)  lattice results. 
These simple models are inspired in the AdS/CFT correspondence and assume that gauge/ gravity duality 
holds in a scenario where conformal invariance is broken through the introduction of an energy scale. 

Another important property of hadrons:  the decay constant,  can also be obtained from these models.  However,  a consistent formulation of an AdS/QCD model that reproduces the  observed behavior of decay constants of vector meson excited states is still lacking. In particular: for radially excited states of heavy vector mesons, the experimental data  lead to decay constants  that decrease with the radial excitation level. 

 We show here that a modified framework of soft wall  AdS/QCD involving an additional dimensionfull parameter,  associated with an ultraviolet energy scale, 
provides  decay constants decreasing with radial excitation level. 
In this version  of the soft wall model the two point function of gauge theory operators  is calculated
at a finite position of the anti-de Sitter space radial coordinate.  
\end{abstract}


\maketitle
\section{ Introduction }  
The discovery of the AdS/CFT correspondence \cite{Maldacena:1997re,Gubser:1998bc,Witten:1998qj} - an exact duality between string theory in certain higher dimensional geometries and supersymmetric $ SU(N)$ conformal gauge theories with large number of colors -  motivated the development of the AdS/QCD phenomenological models.
One of the ideas behind the AdS/QCD approach is to assume  that there is an approximate duality between a field theory living in 
an anti-de Sitter background deformed by the introduction of a dimension-full parameter and a gauge theory where the parameter plays the role of an energy scale.

The idea of associating a cut off in anti-de Sitter space with an infrared energy scale
in the gauge theory appeared in \cite{Polchinski:2001tt}, where  the AdS/CFT duality between a scalar field in anti-de Sitter space and a scalar operator (glueball) in a gauge theory was supplemented with the introduction of a cut off in the radial AdS coordinate. The amplitudes for fixed angle scattering calculated in this way show a ``hard" behavior - they decrease with a power of the energy - as it is observed experimentally for hadrons and is obtained in QCD \cite{Matveev:1973ra,Brodsky:1973kr}.  One remarkable outcome of 
ref. \cite{Polchinski:2001tt},  was to show that,  contrarily to what happens in flat space,  where the amplitudes decrease exponentially with the energy,  string theory in anti-de Sitter space is consistent with the physics of strong interactions, concerning the large energy behavior of fixed angle scattering amplitudes.  

This idea of placing a geometrical cut off in AdS space was then used in \cite{BoschiFilho:2002ta,BoschiFilho:2002vd} as a model for calculating the masses of glueballs. The same approach was then extended to other hadrons and named
as hard wall AdS/QCD model (for a recent review and a wide list of related references, see \cite{Brodsky:2014yha} ). 

The hard wall model fits the available spectrum of radial excitations predicted by lattice 
for scalar glueballs \cite{BoschiFilho:2002ta,BoschiFilho:2002vd} and  experimentally observed for light mesons and baryons \cite{deTeramond:2005su}.  However,  the asymptotic behavior of the spectrum for the hard wall is that the masses grow linearly with the excitation number $n$ for large $n$.  In contrast, the available experimental data  indicate that there is an approximate linear relation between mass squared and radial excitation number. 

Another AdS/QCD model, the soft wall, where  the square of the mass  grow linearly with the
radial excitation number was introduced in \cite{Karch:2006pv}. In this case, the background involves AdS space and a scalar field that acts effectively as a smooth  infrared cut off. 
Many interesting improvements of this model appeared then in the literature. 
In particular, for fermions the soft wall, as formulated in ref. \cite{Karch:2006pv}, does not lead to a discrete spectrum because the dilaton introduced in the action factorizes out in the equations of motion. 
This problem can be overcome, as shown in   
\cite{Vega:2008te,Abidin:2009hr,Gutsche:2011vb}, considering that the  mass of the fermionic field depends on the radial coordinate of AdS space. 
This change in the model is interpreted as effectively 
capturing the dependence of the dimension of the gauge theory operator on the energy scale, as it happens in a non conformal theory. Another important improvement was the discovery that using light-front coordinates, the fith coordinate of AdS space can be interpreted as representing the impact parameter \cite{Brodsky:2007hb,Brodsky:2008pf}. Along the same line, the equivalence between QCD light-front and the equations of both hard and soft wall models was show in ref. \cite{deTeramond:2008ht}.  It is interesting to mention also that the soft wall model was extended to heavy quark hadrons in \cite{Trawinski:2014msa}. 
 
AdS/QCD models provide also a tool for calculating another important property of hadrons: the decay constant. The decay of mesons, that in contrast to baryons are not subjected to a number conservation condition,  is represented 
as a transition from the initial  state to the hadronic vacuum. The amplitude for the decay process is thus calculated from the holographic dual of the transition matrix between the vacuum and the one particle state.
For the soft wall model, expressing  the two point correlator as a sum over transition matrix elements,  one finds the decay constants \cite{Karch:2006pv} (for a detailed discussion see \cite{Grigoryan:2007my}).  The decay constant  $f_n$  of a meson at radial excitation level $n$ is related to the transition matrix between the one particle state and the vacuum.
We use in this article  the definition:  $  \langle 0 \vert \, J_\mu (0)  \,  \vert n \rangle = \epsilon_\mu f_n m_n $.

The results of the soft wall  are degenerate: all the decay constants of the radial excitations of a vector meson are equal. For the hard wall model the decay constants of radial excitations are not degenerate but they  increase with the excitation level.  
 In contrast,  the  experimental results available for heavy vector mesons show that higher excited radial states have smaller decay constants. For the excited states of the light vector meson $\rho$ the available experimental data does not provide an estimate for the decay constants.

For the hard wall model the description of the decay process of vector mesons, as  done in ref. \cite{Grigoryan:2007vg} leads to $ f_n \propto 1/  J_1 ( \gamma_n ) $ where $J_1 $ is the Bessel function of order 
one and $\gamma_n$ are the zeroes of the Bessel function of order zero:  $ J_0 (\gamma_n )  = 0$.   These decay constants  increase with the excitation level since the absolute value of $ J_1 $ evaluated at the zeroes $\gamma_n$     of $J_0$ is a monotonically decreasing function of $n$.      
      So, the hard wall model does not reproduce the observed behavior of the decay constants.

In the original soft wall model there is just  one dimensionfull parameter,  that plays the role of a mass scale.
This infrared parameter is introduced  in the dilaton background of the model and determines the mass spectrum. 
The main motivation of this letter is to propose that the description of decay processes in AdS/QCD should involve an additional dimensionfull parameter, or energy scale.   
In the usual AdS/QCD approach one describes the hadronic decays by means of field  correlators calculated at the AdS boundary. Our proposal is to find the decay constants from the two point correlator calculated at some finite value $ z = z_0$  of the radial coordinate of AdS space.   This corresponds to introducing a new
energy scale: $ 1/z_0 $ in the model.

 A consistent way of calculating field correlators of the gauge theory operators at a finite position $ z = z_0$ of the radial AdS coordinate was presented  in refs. \cite{Afonin:2011ff,Afonin:2012xq}.
 One interesting thing about this procedure   is that the introduction of a second energy scale improves the description of the mass spectra of vector mesons made of heavy quarks.

We specialize our discussion of the soft wall to the case of interest: vector mesons. 
The soft wall action for a vector field  $V_m = (V_\mu,V_z)\,$ ($\mu = 0,1,2,3$), 
assumed to be dual to the gauge theory current $ J^\mu = \bar{q}\gamma^\mu q \,$
reads: 
\begin{equation}
I \,=\, \int d^4x dz \, \sqrt{-g} \,\, e^{- \Phi (z)  } \, \left\{  - \frac{1}{4 g_5^2} F_{mn} F^{mn}
\,  \right\} \,\,, 
\label{vectorfieldaction}
\end{equation}
\noindent where $F_{mn} = \partial_mV_n - \partial_n V_m$ and $\Phi = k^2z^2   $ is the soft wall dilaton background, that plays the role of a smooth infrared cut off and  $k$ is a constant representing the mass scale.  

The  background geometry is anti-de Sitter $AdS_5$ space,  with the metric: 
\begin{equation}
 ds^2 \,\,= \,\, e^{2A(z)}(-dt^2 + d\vec{x}\cdot d\vec{x} + dz^2)\,,
\end{equation}
\noindent where $ A(z) = -log(z/R) $ and the Poincar\'e chart of $AdS$  corresponds to $ (t,\vec{x})\in \mathcal{R}^{1,3},~z\in (0,\infty)$ and $z$ is called radial coordinate.

We use the gauge  $V_z=0$. The boundary value of the remaining components of the vector field:  
$ V^0_{\mu}(x) =
\lim_{z\to 0} V_\mu (x,z) , \mu = 0,1,2,3\, $ are assumed to be, as in the AdS/CFT correspondence,  the sources of the correlation functions of the boundary current operator  $  J^\mu (x) \,\, (= \bar{q}\gamma^\mu q (x) \,)\,$. Namely:
\begin{equation}
 \langle 0 \vert \, J_\mu (x) J_\nu (y) \,  \vert 0 \rangle \, =\, \frac{\delta}{\delta V^{0\mu}(x)} \frac{\delta}{\delta V^{0\nu}(y)}
 \exp  \left( - I_{on shell} \right)\,,
\end{equation}
\noindent where the on shell action is given by the boundary term: 
 \begin{equation}
I_{on \, shell }\,=\, - \frac{1}{2 {\tilde g}_5^2}  \, \int d^4x \,\,\left[  \frac{e^{- k^2 z^ 2  }}{z} V_\mu \partial_z V^ \mu 
\right]_{_{ \! z \to 0 }}
 \,.
\label{onshellaction}
\end{equation}
We have introduced $ {\tilde g}_5^2 =  g_5^2 /R $ that is the relevant dimensionless coupling of the bulk vector field. 
 For comparison with the proposal that we will present in the next section, let us emphasize that this action is calculated at the AdS boundary $z = 0$. 
 One can then write the on shell action in momentum space and decompose the field as 
 \begin{equation} 
 V_\mu (p,z) \,=\, v (p,z) V^0_\mu ( p ) \,,
 \label{Bulktoboundary}
\end{equation}  
 \noindent where    $ v (p,z)   $ is the bulk to boundary propagator that satisfies the equation of motion:
\begin{equation}
\partial_z \Big( \frac{ e^{-k^ 2 z^ 2 }} { z}  \partial_z v (p,z) \Big) + \frac{p^ 2 }{z} e^{-k^ 2 z^ 2 }  v (p,z) \,=\, 0   \,.
\label{BulktoboundaryEOM}
\end{equation}
In order that the factor $ V^0_\mu ( p )$, defined in the decomposition of eq. (\ref{Bulktoboundary}),  works as the source
of the correlators of gauge theory currents, one must impose the boundary condition:  \break $ v (p, z=0) = 1\,$.
On the other hand,  the two point function in momentum space is related to the current-current correlator   by 
 \begin{equation}
  \left( p^2 \eta_{\mu\nu} -  p_\mu p_\nu   \right) \, \Pi ( p^2 ) 
\, =\, \int d^4x \,\, e^{-ip\cdot x} \langle 0 \vert \, J_\mu (x) J_\nu (0) \,  \vert 0 \rangle  \, . 
\label{correlatorand2pointfunction}
\end{equation}  
 Thus the two point function is holographicaly expressed in terms of the bulk to boundary propagator as: 
\begin{equation}
  \Pi ( p^2 )  \, =\,   \frac{1}{{\tilde g}_5^ 2 \, (-p^ 2) } \left[  \frac{ e^{ -k^2 z^2 }  \,  v (p,z) \partial_z v (p,z)  }{  \, z  } 
  \right]_{_{ \! z \to 0 }} \, .  
 \label{hol2point}
\end{equation} 
     
 The two point function has the following spectral decomposition, in terms of masses $m_n$ and   decay constants 
 $f_n$ of the states of the gauge theory:  
 \begin{equation}
\Pi (p^2)  = \sum_{n=1}^\infty \, \frac{f_n^ 2}{(- p^ 2) - m_n^ 2 + i \epsilon} \,. 
\label{2point}
\end{equation} 
In the soft wall model the states are described by the normalizable solutions  $ \phi_n (z)  $   of the equation of motion (\ref{BulktoboundaryEOM}). One can represent the bulk to boundary propagator $ v (p,z)   $ in terms of the normalizable solutions by 
 introducing the Green`s function   
\begin{equation}
  G(p,z,z')  \, =\,  \sum_{n=1}^\infty \frac{  \phi_n (z) \phi_n (z' ) }{( -p^ 2) - m_n^ 2 } \, , 
  \label{Green}
\end{equation}  
 \noindent and writing: 
 \begin{equation}
 v (p,z)  \, =\,  - \left[ \frac{e^{-k^ 2 z^ {\prime 2}} \partial_{z'}   G(p,z',z) }{z' } \right]_{z' \to 0} \,. 
 \label{BulktoboundGreen}
\end{equation}  
The normalized solutions have the form:  $ \phi_n(z) = \sqrt{ 2(n-1)!/n!} (kz)^{2}L_{n-1}^1(k^2z^2)\,$,
where  $L_n^1(x)$ are the generalized Laguerre polynomials of order one. They satisfy the boundary condition 
$\lim_{z\to 0}  \phi_n (z) = 0$ and correspond to solutions of eq. (\ref{BulktoboundaryEOM}) with momenta:
$ - p_n^2 = m_n^2 = 4k^2 n \, ,~~n=1,2,...  \,$.

The bulk to boundary propagator defined in eq. (\ref{BulktoboundGreen})   satisfies the equation of motion 
 (\ref{BulktoboundaryEOM}) and also the boundary condition $ v(p, 0) = 1$. It is important to discuss the subtle way in which this
 boundary condition is satisfied.   
 If one simply takes the definition for the bulk to boundary propagator from the Green's function one finds an infinite sum 
 of terms with the propagator poles $1/(-p^ 2 - m^ 2 ) $ that would in principle lead to a dependence of $ v(p, 0) $ on the momentum 
 $p$. However, the boundary condition $\lim_{z\to 0}  \phi_n (z) = 0$ leads to a sum of singular terms with vanishing coefficients, 
 that needs to be regularized.  Such a regularization is presented in ref. \cite{Grigoryan:2007my},  where an integral representation for the bulk to boundary propagator is found. This integral representation is well defined in the interval $0 \le z $. It is explicitely equal to one in $z=0$ and coincides with the expression (\ref{BulktoboundGreen})  when $z > 0$.

 The two point function obtained holographically in eq. (\ref{hol2point})  has the same pole structure of eq.  (\ref{2point}) , with the residues corresponding to the decay constants. Thus:
\begin{equation}
 f_n    \, =\,  \left[ \frac{e^{-k^ 2 z^ 2 }\,\partial_z \phi_n (z) }{ m_n \, {\tilde g}_5 \, z}  \right]_{ z \to 0 }  \,=\, 
 \frac{  k \sqrt{ 2} }{{\tilde g}_5}  \,.
 \label{decayconstants}
\end{equation} 
 
\noindent This result shows that in the soft wall model all the radial excitations of a  vector meson
have the same decay constant. This degeneracy contrasts with the decrease with excitation level $n$  in the decay constants 
obtained from experimental data for heavy vector mesons.

This problem was discussed in reference \cite{Grigoryan:2010pj}, where a finite temperature AdS/QCD model for quarkonium in a plasma was discussed. In order to describe the finite temperature behavior, the authors proposed a 
procedure to fit the zero temperature parameters.
This was done by adding extra terms in the equation of motion for the normalizable modes, in such a way that one arrives at a model with four parameters
that are fixed by the four quantities available: the masses and decay constants
of $J /\psi $ and $\psi^\prime$.   

\section{Alternative description of the decay process}

The soft wall model, as reviewed in the previous section,  contains one dimensionfull parameter:  $k$,  introduced in the dilaton background. This parameter plays the role of an infrared (or mass)  scale of the model and determines the spectrum of the 
vector mesons: $ m_n^2 = 4k^2 n $ as well as the decay constants.  The linear relation between the square of the mass and the 
excitation level is the expected asymptotic  behavior.  
However, the observed behavior of decay constants of vector mesons is that they decrease with the excitation level
while the soft wall model predicts degenerate values. 

The relation between the vacuum expectation value of the product of two gauge currents and the decay constants
 \begin{equation}
  \int d^4x \,\, e^{-ip\cdot x} \langle 0 \vert \, J_\mu (x) J_\nu (0) \,  \vert 0 \rangle  \,= \,
   \left( p^2 \eta_{\mu\nu} -  p_\mu p_\nu   \right) \,   \sum_{n=1}^\infty \, \frac{f_n^ 2}{(- p^ 2) - m_n^ 2 + i \epsilon} \,,
\label{2pointdifferent}
\end{equation} 
comes from introducing on the lefthand side of this equation a basis of the states of the theory. The decay constants are related to the transition matrices between the one particle states and the vacuum as:  $  \langle 0 \vert \, J_\mu (0)  \,  \vert n \rangle = \epsilon_\mu f_n m_n $. 
Our point of view is that, in order to incorporate in the model the effect of the interactions that govern the transition processes 
from the one particle state to the vacuum, corresponding to the hadron decay,   one needs to introduce one additional dimensionfull parameter in the soft wall.  This can be done by calculating holographicaly the operator product of currents of eq. (\ref{2pointdifferent})  at a finite  location $ z = z_0$ of the radial coordinate.  The new parameter, namely $ 1/z_0 $,  corresponds to an energy scale  where the transition matrices are calculated. 

The idea of  introducing an ultraviolet cutoff in the soft wall model was previously considered in ref. \cite{Evans:2006ea}.  In this reference  normalized solutions ${\tilde \phi}_n $ satisfying   boundary condition
at a position corresponding, in our notation, to a minimum value of the z coordinate $z =  1/ \Lambda $
were studied numerically. The condition imposed on the solutions was $ \lim_{z\to 1/\Lambda} {\tilde \phi}_n (z) \sim \, z^ 2 \,$ ;
$\,\, \lim_{z\to 1/\Lambda} {\tilde \phi}^{\prime}_n (z) \sim \, 2 z  \,$. 
The spectrum of masses obtained for the $\rho$ meson from this approach showed a better fit to the experimental data, in comparison with the original soft wall model. 

In this reference \cite{Evans:2006ea}  the decay constants were also discussed. It was assumed that with the introduction of the UV cutoff they should be given (using our choice of radial coordinate and our convention for decay constants)  by:

\begin{equation}
 f_n    \,    =    \,  \left[ \frac{ \partial_z {\tilde \phi}_n (z) }{m_n \,  g_5 \, z}  \right]_{ z \to 1/\Lambda } \,.
 \label{EvansDecay}
\end{equation} 

Using this relation, a decrease in the decay constants with the excitation level was found. However there is a problem with expression  (\ref{EvansDecay}). The boundary value of the bulk field plays the role of the source of the gauge current correlators. When one takes the boundary to be at some finite value of the coordinate $ z = 1/\Lambda $ and decompose the field, as in eq. (\ref{Bulktoboundary}),
$ V_\mu (p,z) \,=\, v (p,z) V^0_\mu ( p ) \, $,  the source is  $ V^0_\mu ( p ) = \lim_{ z \to 1/\Lambda }  V_\mu (p,z) $
and the bulk to boundary propagator has to satisfy the condition: $ \lim_{ z \to 1/\Lambda } v (p,z) = 1$.
In order to represent $v(p,z) $  in terms of the normalized solutions $ {\tilde \phi}_n $ in the same way as in equations (\ref{Green}) and (\ref{BulktoboundGreen})  it is necessary that the condition $ \lim_{z\to 1/\Lambda} {\tilde \phi}_n (z) = 0 \,$,
holds for the normalized solutions. This way, one finds for the bulk to boundary propagator at $ z = 1/\Lambda $  
a sum of singular terms with vanishing coefficients  that  can be regularized to unity, as discussed in section II.

For a non vanishing boundary condition,  like the one used in ref. \cite{Evans:2006ea},   
the bulk to boundary propagator will be a sum of propagator pole terms with finite coefficients, keeping a momentum dependence. So, the condition: $v(p, 1/\Lambda ) = 1$ is not satisfied and $ V^0_\mu ( p )$ is not the source
of the correlators. Thus, in this case, one can not assume the expression (\ref{EvansDecay})
to represent the decay constants.

What comes out as a consistent approach to describe the decay constants by calculating the field correlators at a finite position $ z = z_0$  of the radial coordinate is to follow a procedure presented in  
refs. \cite{Afonin:2011ff,Afonin:2012xq}.  
 In these references a cut off in the radial AdS coordinate was introduced so as to play the role of an ultraviolet cut off in the gauge theory. We will relate the same sort of cut off with the decay process. The bulk to boundary propagator is written as a solution of the equation of motion (\ref{BulktoboundaryEOM})  but now excluding the region $0 < z < z_0$. The solution is divided by a constant (in the $z$ coordinate) so as to  satisfy by construction the boundary condition $v(p,z_0) = 1$. Namely: 
 \begin{equation} 
 v (p,z ) \, = \, \frac{ U (p^ 2/ 4k^ 2 , 0, k^2 z^ 2 ) }{U (p^ 2/ 4k^ 2 , 0, k^2 z_0^ 2 )}\,,
 \label{bulktoboundary2}
\end{equation} 
 
\noindent where $U(a,b,c)$ is the Tricomi function. Using   the new on shell action given by:
 \begin{equation}
I_{on \, shell }\,=\, - \frac{1}{2 {\tilde g}_5^2}  \, \int d^4x \,\,\, \frac{e^{- k^2 z^ 2  }}{z} V_\mu \partial_z V^ \mu 
{\Big \vert}_{_{ \! z \to z_0 }} 
 \,,
\label{onshellaction2}
\end{equation}

\noindent we calculate the two point function:
\begin{equation}
  \Pi ( p^2 )  \, =\,   \frac{1}{{\tilde g}_5^ 2 \, (-p^ 2) }  \frac{ e^{ -k^2 z^2 }  \,  v (p,z) \partial_z v (p,z)  }{  \, z  } 
  {\Big \vert}_{_{ \! z \to z_0 }} 
   \, .
 \label{hol2pointnew}
\end{equation} 

\noindent     Then, using the new bulk to boundary propagator of eq. (\ref{bulktoboundary2})  we get:
 \begin{equation} 
 \Pi ( p^ 2)  \, = \, \frac{1}{2 {\tilde g}_5^ 2 }\frac{ e^ {-k^ 2 z_0^ 2} U (1 + p^ 2/ 4k^ 2 , 1, k^2 z_0^ 2 ) }{U (p^ 2/ 4k^ 2 , 0, k^2 z_0^ 2 )}\,,
 \label{correlator2}
\end{equation} 

\noindent where a recursion relation for the Tricomi functions has been used. 

The spectrum of masses and decay constants comes from the analysis of the poles in the momentum variable $p^2$.
The Tricomi function in the numerator is non singular in $p^ 2$ , so the two point function of eq. (\ref{bulktoboundary2}) is only singular at the zeros of the denominator. 
Furthermore, the  poles $p^ 2_n$ of the function  $\Pi ( p^ 2)$  are simple. So, in a neighborhood of $ p^ 2 = p^ 2_n$ one can approximate  
the two point function by:
\begin{equation}
\lim_{p^ 2 \to p^ 2_n} \Pi (p^2)   \approx \frac{f_n^ 2}{(-p^ 2)  + p_n^ 2 } \,.
\label{2pointlimit}
\end{equation} 
\noindent We associate the coefficients of the approximate expansion near the pole with the decay constant $f_n$ in analogy with the exact expansion shown in eq. (\ref{2point}). This way we get the masses from the localization of the poles of the two point function and the decay constants from the
corresponding  coefficient. More precisely, let $\chi_n$ be the roots of the Tricomi function: 
\begin{equation}
 U (  \chi_n \,  , 0, k^2 z_0^ 2 ) = 0 \,,
 \label{roots}
 \end{equation} 
 \noindent then the holographic vector meson masses are: $ m^2_n \, =  \, 4 k^ 2  \, \chi_n \,$.
  The decay constants are calculated numerically from the fit to the approximate form presented in equation (\ref{2pointlimit}).  That means:
\begin{equation}
f_n^ 2 \,=\, 
\lim_{p^2 \to p^2_n} \Big( (-p^ 2)  + p_n^ 2 \Big) \, 
\Pi (p^2)   \ 
 \,.
\label{numerical decay}
\end{equation} 
The coupling ${\tilde g}_5 = g_5 /\sqrt{R} $ of the vector field in the AdS bulk can be obtained by comparison with QCD, as explained in references  \cite{Karch:2006pv,Grigoryan:2007my}, wich gives:
${\tilde g}_5 \, =\, 2 \pi $.

\section{Confronting the model with the available data } 

\begin{table}[h]
\centering
\begin{tabular}[c]{|c||c||c||c|}
\hline 
\multicolumn{4}{|c|}{   Charmonium  data  } \\
\hline
 & Masses (MeV)   & $\Gamma_{V \to e^+e^-} $ (keV) & Decay constants (MeV)  \\
\hline
$ \,\,\,\, 1S  \,\,\,\,$ & $3096.916 \pm 0.011 $ & $ \,\,\,\, 5.547\pm 0.14  \,\,\,\,$ & $ 416.2 \pm 5.3 $ \\ 
\hline
$ \,\,\,\,   2S  \,\,\,\,$ & $ 3686.109 \pm 0.012 $ & $\,\,\,\,  2.359 \pm 0.04\,\,\,\,$  & $ 296.1 \pm 2.5$ \\
\hline 
$ \,\,\,\,3S \,\,\,\,$ & $ 4039 \pm 1 $ &$  \,\,\,\, 0.86\pm 0.07 \,\,\,\,$ & $ 187.1  \pm 7.6 $\\ 
\hline
\,\,\,\, $ 4S$  \,\,\,\,&$ 4421 \pm 4 $ &$ \,\,\,\, 0.58\pm 0.07 \,\,\,\, $ & $160.8  \pm 9.7$ \\
\hline
 \end{tabular}
 \caption{Experimental masses and electron-positron widths from \cite{Agashe:2014kda} and the corresponding decay constants for the Charmonium S-wave resonances.  }
\end{table}

\begin{table}[h]
\centering
\begin{tabular}[c]{|c||c||c||c|}
\hline 
\multicolumn{4}{|c|}{  Bottomonium  data   } \\
\hline
 &  Masses (MeV)   & $ \Gamma_{V \to e^+e^-} $ (keV) &   Decay constants (MeV) \\
\hline
$\,\,\,\, 1S \,\,\,\,$ & $ 9460.3\pm 0.26 $ &$ \,\,\,\, 1.340 \pm 0.018 \,\,\,\, $ & $ 715.0 \pm 2.4 $ \\ 
\hline
$\,\,\,\, 2S \,\,\,\,$ & $ 10023.26 \pm 0.32 $ & $ \,\,\,\,  0.612 \pm 0.011\,\,\,\,$  & $ 497.4 \pm 2.2 $  \\
\hline 
$\,\,\,\,3S \,\,\,\,$ & $ 10355.2 \pm 0.5 $ & $ \,\,\,\, 0.443 \pm 0.008 \,\,\,\, $ & $ 430.1  \pm 1.9 $ \\ 
\hline
$ \,\,\,\, 4S  \,\,\,\,$ & $ 10579.4 \pm 1.2 $ & $ \,\,\,\, 0.272 \pm 0.029 \,\,\,\,$  & $ 340.7  \pm 9.1 $ \\
\hline
\end{tabular}   
\caption{Experimental masses and electron-positron widths from \cite{Agashe:2014kda} and the corresponding decay constants for the Bottomonium S-wave resonances.  }
\end{table}

The experimental values for the masses $m_{_V} $  and the electron positron decay widths  
$\Gamma_{ V \to e^+ e^ - } $ for  vector mesons are taken from ref. \cite{Agashe:2014kda}. 
The decay constant of a vector meson state is related to the corresponding mass and width by \cite{Hwang:1997ie}:
\begin{equation} 
f_{_V}^ 2 \,=\, \frac{3 m_{_V} \Gamma_{ V \to e^+ e^ - }  }{4 \pi \alpha^ 2 c_{_V}}\,,
 \label{decay-widths}
 \end{equation} 
 \noindent where $\alpha = 1/137 $ and $c_{_V} $ for the two families that we consider are  $  c_{J/\Psi}  = 4/9 $ and $c_{\Upsilon} = 1/9$.   
 The uncertainties in the decay constants are: 
\begin{equation} 
\delta f_{_V}  \,=\, \frac{3  }{8 \pi \alpha^ 2 c_{_V}  f_{_V} }  \left(  m_{_V} \delta \Gamma_{ V \to e^+ e^ - }   + \Gamma_{ V \to e^+ e^ - }   \delta  m_{_V}   \right)\,.
 \label{decay-errors}
 \end{equation}

  In table {\bf I} we show the experimental values of the  masses and  widths  and the associated decay constants with the corresponding uncertainties for the charmonium vector meson  $J/\psi$, made of  a charmed quark anti-quark pair and  for the first three radially excited S-wave resonances.  This table shows that the decay constants decrease monotonically with the radial excitation level. 
  
  Then, in table {\bf II} we show the experimental data and the associated decay constants with the corresponding uncertainties for the botomonium vector meson  $ \Upsilon $, made of  a bottom quark anti-quark pair and  for the first three radially excited S-wave resonances.  As it happens for the charmonium, the decay constants decrease monotonically with the radial excitation level.

Now we present the results for the masses,  obtained from equations (\ref{roots})  and 
the decay constants obtained from equation (\ref{numerical decay}) for charmonium and bottomonium S-wave states. 
 A nice fit to the experimental data is obtained  using the parameters 
 \begin{equation}
  k_c = 1.2 GeV ; \,  k_b = 3.4 GeV ; \, 1/z_0 = 12.5 GeV,  
  \label{parameters}
  \end{equation}   
 \noindent where $ k_c$ and $k_b$ are the values of the constants $k$ used for charmonium and bottomonium, respectively. The values are different, reflecting the difference in the masses of charm and bottom quarks. The energy scale $1/z_0$ is taken 
 to be flavor independent and represents a characteristic energy of the decay process. In the non hadronic decay a very heavy vector meson annihilates into light leptons. So, the ultraviolet scale is of order of the heavier masses that we consider.    
\begin{table}[h]
\parbox{.45\linewidth}{
\centering
\begin{tabular}[c]{|c||c||c|} 
\hline 
\multicolumn{3}{|c|}{  Results for Charmonium states   } \\
\hline
 &  Masses    &   Decay constants  \\
 \hline
$\,\,\,\, 1S \,\,\,\, $ &  $\,\,\,\,2410 \,\, \,\,\,\, $ &    $\,\,\,\, 258.8  \,\,\,\, $\\ 
\hline
$ \,\,\,\, 2S \,\,\,\, $ & $ 3409  \,\,  $ & $251.7  $ \\
\hline 
$ \,\,\,\, 3S  \,\,\,\, $  & $4174  $  & $ 245.9 $ \\ 
\hline
$\,\,\,\, 4S  \,\,\,\, $ & $ 4819$ & $ 241.0  $ \\
\hline
\end{tabular} 
\caption{  Masses   and decay constants in MeV obtained for the Charmonium states.       }
 }
\hfill
\parbox{.45\linewidth}{       
 \centering   
\begin{tabular}[c]{|c||c||c|} 
\hline 
\multicolumn{3}{|c|}{  Results for Bottomonium states } \\
\hline
 &  Masses   &   Decay constants \\
 \hline
$\,\,\,\, 1S \,\,\,\, $ & $ \,\,\,\, 7011   \,\,\,\, $ & $ \,\,\,\, 627 \,\,\,\,$  \\
\hline
$\,\,\,\, 2S \,\,\,\, $ & $9883  $  & $ 574  $ \\
\hline 
$ \,\,\,\, 3S  \,\,\,\, $  & $ 12077 $  & $ 538  $\\ 
\hline
$ \,\,\,\, 4S  \,\,\,\, $ & $ 13923 $   &  $ 512 $ \\
\hline
\end{tabular}   
\caption{  Masses   and decay constants obtained for the Bottomonium S-wave resonances.     }
       }
\end{table} 
We show  in tables  {\bf III }  and {\bf IV}  the results obtained for the charmonium and bottomonium  S-wave states, respectively.  One can see that the model produces a decrease of the  decay constants with the radial excitation level, as it happens  with the available data for these particles.  

 One interesting way to characterize the model in terms of predictability is to
 define the rms error for estimating $N$ quantities using a model with $N_p$ parameters as:
 \begin{equation}
 \delta_{rms} = \sqrt{ \frac{1}{(N - N_ p )}\sum_i^N  \left( \frac{\delta O_i}{O_i} \right)^ 2 }
 \label{error}
 \end{equation}
 \noindent where $O_i$ is the average experimental value and $\delta O_i$ is the deviation of the value provided 
 by the model. 
 Considering all the 16 results of tables  {\bf III }  and {\bf IV} , obtained using three free parameters: 
  $ k_b, k_c $ and $ 1/z_0 $ we find $  \delta_{rms}  = 30 \, \% $.   
  It is important to remark that we are fitting two different properties, the mass and the decay constant, of four  resonances of two different hadrons using only three parameters in the model. 
  So, the error of 30 $\% $  is reasonable, considering the simplicity of the model and the complexity of the strong coupling physics involved in the heavy meson structure.

\section{  Final Comments }

The experimental data for the first four S-wave resonances of charmonium and bottomonium 
lead to decay constants that exhibit a clear decrease with the radial excitation level.
In this article we proposed a consistent AdS/QCD approach where the decay constants are calculated  from two point operator products at a finite value of the radial coordinate of anti-de Sitter space.
The model was formulated with three independent parameters and was applied to the calculation of sixteen  
independent observables: eight  masses and eight decay constants. 

A consistent AdS/QCD  method for finding decay constants that decrease with radial excitation was lacking in the literature. This property is reproduced in our model, in contrast to the original soft wall, providing a more realistic description of heavy vector mesons.

\bigskip
  
\noindent {\bf Acknowledgments:}  We thank J.M. R. Rold\'{a}n for important discussions and Romulo Rodrigues da Silva for important correspondence. 
N.B. and S.D. are partially supported by CNPq and M.A.M. is supported by Vicerrectoria de Investigaciones de La Universidad
de los Andes.


\begin{thebibliography}{ABC}

\bibitem{Maldacena:1997re}
  J.~M.~Maldacena,
  Adv.\ Theor.\ Math.\ Phys.\  {\bf 2}, 231 (1998)
  [Int.\ J.\ Theor.\ Phys.\  {\bf 38}, 1113 (1999)].
 [arXiv:hep-th/9711200].

\bibitem{Gubser:1998bc}
  S.~S.~Gubser, I.~R.~Klebanov and A.~M.~Polyakov,
  Phys.\ Lett.\  B {\bf 428}, 105 (1998).
  [arXiv:hep-th/9802109].

\bibitem{Witten:1998qj}
  E.~Witten,
  Adv.\ Theor.\ Math.\ Phys.\  {\bf 2}, 253 (1998).
  [arXiv:hep-th/9802150].

\bibitem{Polchinski:2001tt}
  J.~Polchinski and M.~J.~Strassler,
  Phys.\ Rev.\ Lett.\  {\bf 88}, 031601 (2002)
  [arXiv:hep-th/0109174].

\bibitem{Matveev:1973ra}
  V.~A.~Matveev, R.~M.~Muradian and A.~N.~Tavkhelidze,
  Lett.\ Nuovo Cim.\  {\bf 7}, 719 (1973).
 
\bibitem{Brodsky:1973kr}
  S.~J.~Brodsky and G.~R.~Farrar,
  Phys.\ Rev.\ Lett.\  {\bf 31}, 1153 (1973);
  Phys.\ Rev.\  D {\bf 11}, 1309 (1975).


\bibitem{BoschiFilho:2002ta}
  H.~Boschi-Filho and N.~R.~F.~Braga,
  Eur.\ Phys.\ J.\  C {\bf 32}, 529 (2004)
  [arXiv:hep-th/0209080].
  
\bibitem{BoschiFilho:2002vd}
  H.~Boschi-Filho and N.~R.~F.~Braga,
  JHEP {\bf 0305}, 009 (2003)
  [arXiv:hep-th/0212207].

\bibitem{Brodsky:2014yha} 
  S.~J.~Brodsky, G.~F.~de Teramond, H.~G.~Dosch and J.~Erlich,
  Phys.\ Rept.\  {\bf 584}, 1 (2015)
  doi:10.1016/j.physrep.2015.05.001
  [arXiv:1407.8131 [hep-ph]].


\bibitem{deTeramond:2005su} 
  G.~F.~de Teramond and S.~J.~Brodsky,
  Phys.\ Rev.\ Lett.\  {\bf 94}, 201601 (2005)
  [hep-th/0501022].
 
 
\bibitem{Karch:2006pv}
  A.~Karch, E.~Katz, D.~T.~Son and M.~A.~Stephanov,
  Phys.\ Rev.\  D {\bf 74}, 015005 (2006)
  [arXiv:hep-ph/0602229].
  
 
\bibitem{Vega:2008te}
  A.~Vega, I.~Schmidt,
  Phys.\ Rev.\  {\bf D79}, 055003 (2009).
  [arXiv:0811.4638 [hep-ph]].
  
\bibitem{Abidin:2009hr}
  Z.~Abidin and C.~E.~Carlson,
  Phys.\ Rev.\  D {\bf 79}, 115003 (2009)
  [arXiv:0903.4818 [hep-ph]].

\bibitem{Gutsche:2011vb} 
  T.~Gutsche, V.~E.~Lyubovitskij, I.~Schmidt and A.~Vega,
  Phys.\ Rev.\ D {\bf 85}, 076003 (2012)
  [arXiv:1108.0346 [hep-ph]].

\bibitem{Brodsky:2007hb} 
  S.~J.~Brodsky and G.~F.~de Teramond,
  Phys.\ Rev.\ D {\bf 77}, 056007 (2008)
  doi:10.1103/PhysRevD.77.056007
  [arXiv:0707.3859 [hep-ph]].
  
\bibitem{Brodsky:2008pf} 
  S.~J.~Brodsky and G.~F.~de Teramond,
  Phys.\ Rev.\ D {\bf 78}, 025032 (2008)
  doi:10.1103/PhysRevD.78.025032
  [arXiv:0804.0452 [hep-ph]].
  
\bibitem{deTeramond:2008ht} 
  G.~F.~de Teramond and S.~J.~Brodsky,
  Phys.\ Rev.\ Lett.\  {\bf 102}, 081601 (2009)
  doi:10.1103/PhysRevLett.102.081601
  [arXiv:0809.4899 [hep-ph]].
  
\bibitem{Trawinski:2014msa} 
  A.~P.~Trawiński, S.~D.~Głazek, S.~J.~Brodsky, G.~F.~de Téramond and H.~G.~Dosch,
  Phys.\ Rev.\ D {\bf 90}, no. 7, 074017 (2014)
  doi:10.1103/PhysRevD.90.074017
  [arXiv:1403.5651 [hep-ph]].
  
     
\bibitem{Grigoryan:2007my} 
  H.~R.~Grigoryan and A.~V.~Radyushkin,
  Phys.\ Rev.\ D {\bf 76}, 095007 (2007)
  [arXiv:0706.1543 [hep-ph]].
  
       
\bibitem{Grigoryan:2007vg} 
  H.~R.~Grigoryan and A.~V.~Radyushkin,
  Phys.\ Lett.\ B {\bf 650}, 421 (2007)
  [hep-ph/0703069].  
  
  
\bibitem{Afonin:2011ff} 
  S.~S.~Afonin,
  Phys.\ Rev.\ C {\bf 83}, 048202 (2011)
  [arXiv:1102.0156 [hep-ph]].

\bibitem{Afonin:2012xq} 
  S.~S.~Afonin,
  Int.\ J.\ Mod.\ Phys.\ A {\bf 27}, 1250171 (2012)
  [arXiv:1207.2644 [hep-ph]].

 
\bibitem{Grigoryan:2010pj} 
  H.~R.~Grigoryan, P.~M.~Hohler and M.~A.~Stephanov,
  Phys.\ Rev.\ D {\bf 82}, 026005 (2010)
  [arXiv:1003.1138 [hep-ph]].

\bibitem{Evans:2006ea} 
  N.~Evans and A.~Tedder,
  Phys.\ Lett.\ B {\bf 642}, 546 (2006)
  [hep-ph/0609112].

\bibitem{Agashe:2014kda} 
  K.~A.~Olive {\it et al.} [Particle Data Group Collaboration],
  Chin.\ Phys.\ C {\bf 38}, 090001 (2014).
     
\bibitem{Hwang:1997ie} 
  D.~S.~Hwang and G.~H.~Kim,
  Z.\ Phys.\ C {\bf 76}, 107 (1997)
  [hep-ph/9703364].     
  
   

 

\end{thebibliography}
 \end{document}